\documentclass{article}
\usepackage{jamaica04}
\usepackage{graphicx}
\frompage{000} \topage{000}                                              
\def\rightmark{Fluctuations from Thermalization }
\def\leftmark{S. Gavin}

\title{Fluctuations from Thermalization at RHIC}
\authors{
{Sean Gavin}\\[2.812mm]
{\normalsize
Department of Physics and Astronomy, Wayne State University \\
Detroit, MI 48201, USA
}}

\abstract{The centrality dependence of dynamic fluctuations of the
transverse momentum and the net charge can signal the approach to
local thermal equilibrium in nuclear collisions. I explore this
signal by comparing transport-theory calculations to STAR and
PHENIX data at a range of energies. In particular, I find that
this model can describe PHENIX data on the dependence of
fluctuations on the $p_t$ range in which they are measured.}


\begin{document}

\maketitle
\setcounter{page}{1}

\section{Introduction}


Measurements of the event-by-event fluctuations of the mean
transverse momentum exhibit substantial dynamic contributions
\cite{Mitchell}. PHENIX and STAR data for Au+Au collisions show
that such fluctuations increase as centrality increases
\cite{PHENIX,Pruneau}. In addition, the measured dynamic
net-charge fluctuations may show an analogous behavior
\cite{Pruneau}. Importantly, the event-averaged transverse
momentum $\langle p_t\rangle$ -- a quantity unaffected by
fluctuations -- exhibits a strikingly similar behavior
\cite{StarMean,Adler:2003cb}.

I ask whether the onset of local thermal equilibrium can explain
the common centrality dependence of these quantities. I seek a
nonequilibrium explanation because the strongest observed changes
occur in peripheral collisions, where equilibrium is not expected.
This is important because similar behavior can accompany the
hadronization of quark-gluon plasma, but only if the matter is
thermalized \cite{Wilk}.  My discussion follows the
transport-theory formulation in ref.~\cite{Gavin04}. Here, I
present work in progress confronting new measurements of the
variation of $p_t$ fluctuations as a function of the observed
$p_t$ region \cite{PHENIX}; see fig.~\ref{fig:ptMax}.

Dynamic fluctuations are generally determined from the measured
fluctuations by subtracting the statistical value expected, e.g.,
in equilibrium \cite{PruneauGavinVoloshin}. For particles of
momenta $\mathbf{p}_1$ and $\mathbf{p}_2$, dynamic multiplicity
fluctuations are characterized by the robust variance
\begin{equation}\label{eq:DynamicMult}
    R_{AA}={{\langle N^2\rangle -\langle N\rangle^2 -\langle N\rangle}\over{\langle
    N\rangle^2}}={{1}\over{\langle N\rangle^{2}}}\int\! d\mathbf{p}_{1}d\mathbf{p}_{2}\,
    r(\mathbf{p}_{1},\mathbf{p}_{2}),
\end{equation}
where $\langle \cdots\rangle$ is the event average. This quantity
depends only on the two-body correlation function
$r(\mathbf{p}_{1},\mathbf{p}_{2}) =
N(\mathbf{p}_{1},\mathbf{p}_{2}) -
N(\mathbf{p}_1)N(\mathbf{p}_2)$, see also \cite{Wilk01}. It is
obtained from the multiplicity variance by subtracting its Poisson
value $\langle N\rangle$. To make $R_{AA}$ robust, we divide by
$\langle N\rangle^2$ to minimized the effect of experimental
efficiency \cite{PruneauGavinVoloshin}. For dynamic $p_t$
fluctuations one similarly finds
\begin{equation}\label{eq:Dynamic}
    \langle \delta p_{t1}\delta p_{t2}\rangle =
    \int\! d\mathbf{p}_{1}d\mathbf{p}_{2}\,
    {{r(\mathbf{p}_{1},\mathbf{p}_{2})}\over{\langle
    N(N-1)\rangle}}
    \delta p_{t1} \delta p_{t2},
\end{equation}
where $\delta p_{ti} = p_{ti}-\langle p_t\rangle$; STAR measures
this observable. The observable measured by PHENIX satisfies
$F_{p_t} \approx N \langle\delta p_{t1}\delta
p_{t2}\rangle/2\sigma^2$ when dynamic fluctuations are small
compared to statistical fluctuations $\sigma^2 = \langle
p_t^2\rangle - \langle p_t\rangle^2$.


Thermalization occurs as scattering drives the phase space
distribution within a small fluid cell toward a Boltzmann
distribution, which varies in space through the temperature
$T(\mathbf{x}, t)$. The time scale for this process is the
relaxation time $\nu^{-1}$. In contrast, density differences
\emph{between} fluid cells must be dispersed by transport from
cell to cell. The time needed for diffusion to disperse a dense
fluid ``clump'' of size $L \sim (|\nabla n|/n)^{-1}$ is $t_{\rm
d}\sim \nu L^2/v_{\rm th}^2$, where $v_{\rm th}\sim 1$ is the
thermal speed of particles. This time can be much larger than
$\nu^{-1}$ for a sufficiently large clumps. Global equilibrium, in
which the system is uniform, can be only obtained for $t \gg
t_{\rm d}$. However, the rapid expansion of the collision system
prevents inhomogeneity from being dispersed prior to freeze out.

Dynamic fluctuations depend on the number of independent particle
``sources.'' This number changes as the system evolves. Initially,
these sources are the independent strings formed as the nuclei
collide. The system is highly correlated along the string, but is
initially uncorrelated in the transverse plane. As local
equilibration proceeds, the clumps become the sources. The
fluctuations at this stage depend on number of clumps as
determined by the clump size, i.e., the correlation length in the
fluid. Dynamic fluctuations would eventually vanish if the system
were to reach global equilibrium, where statistical fluctuations
would be determined by the total number of particles.

\section{Mean $p_t$ and its Fluctuations}

\begin{figure}
\begin{center}
\includegraphics[width=2.4in]{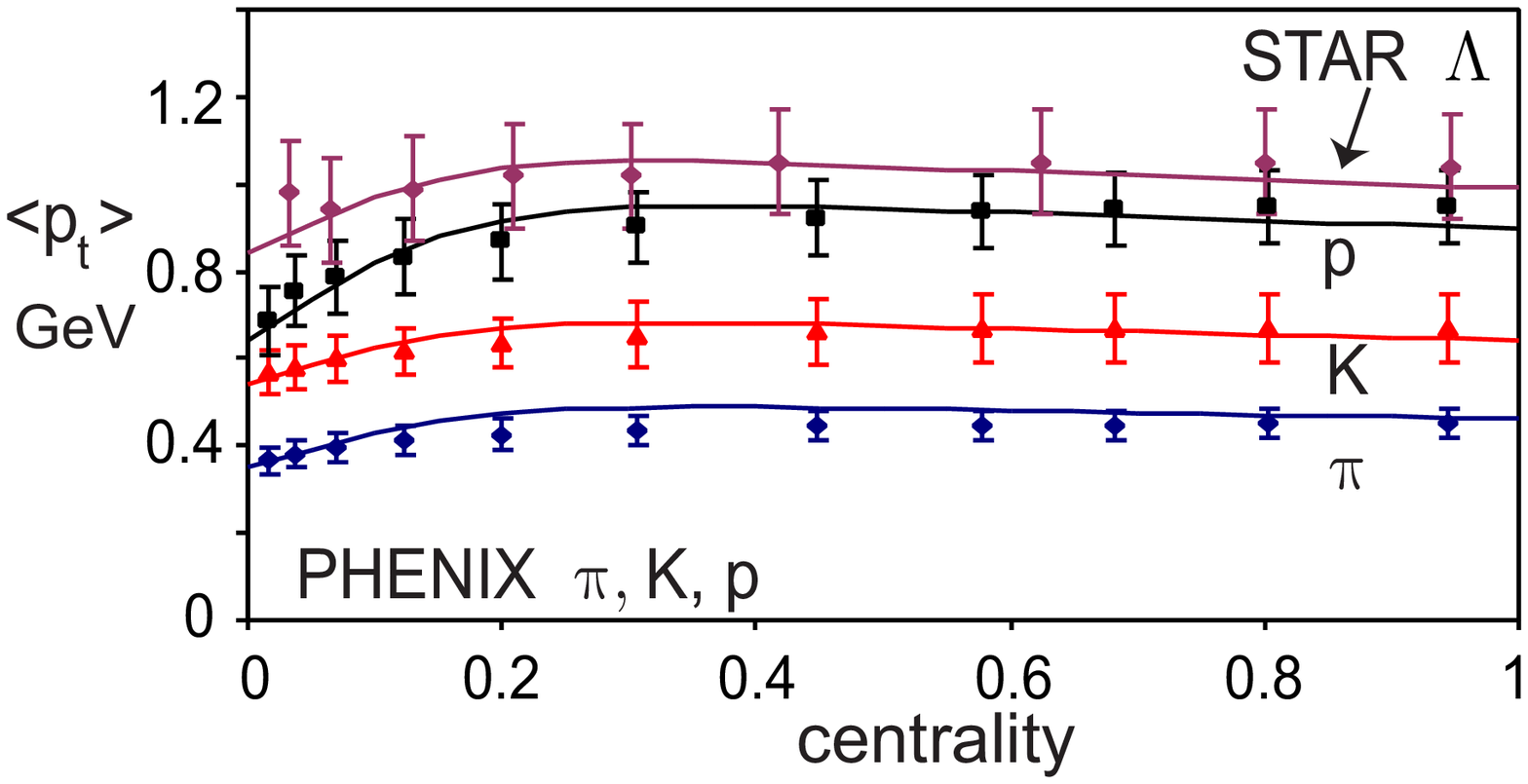}
\includegraphics[width=2.5in]{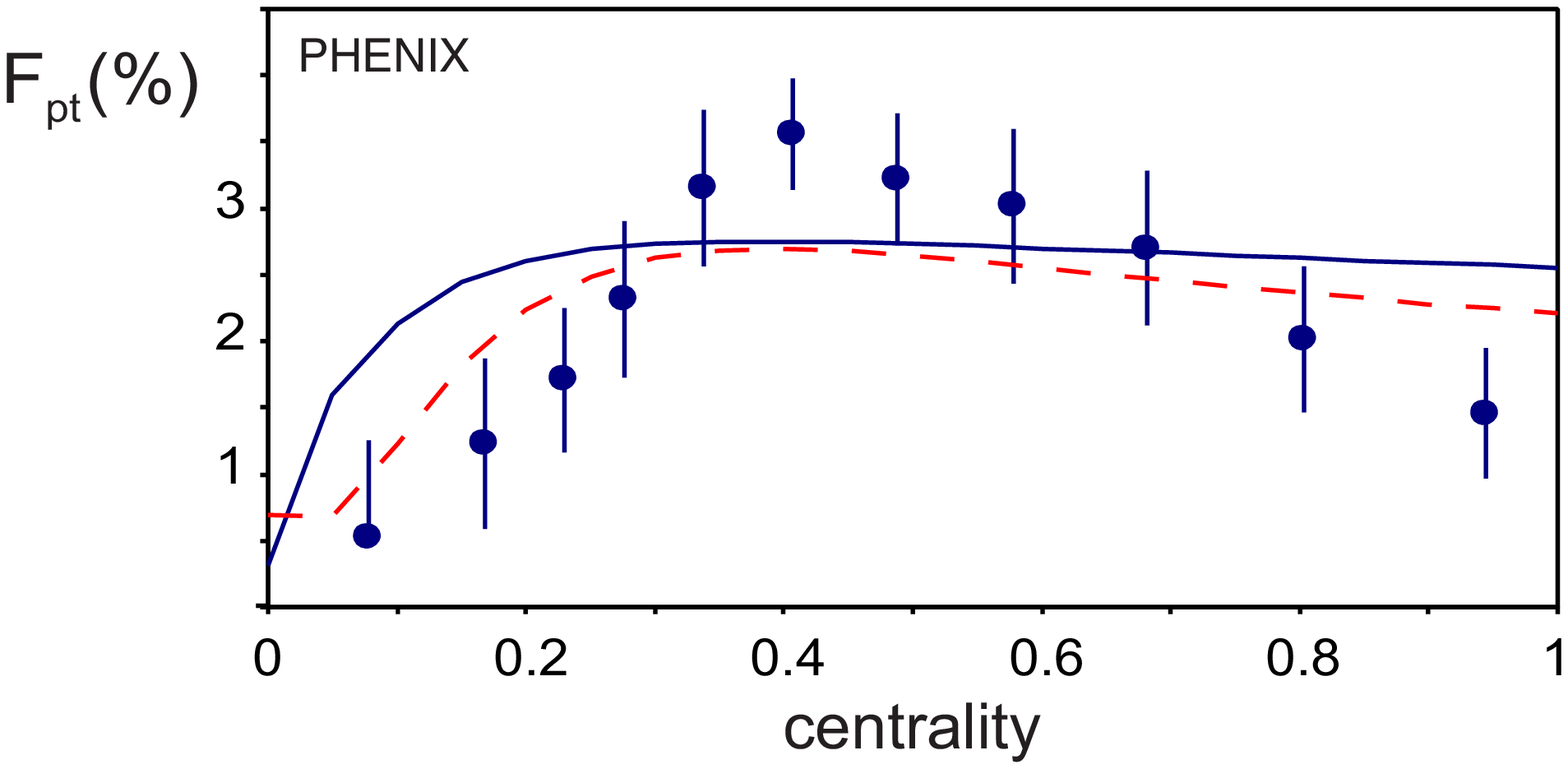}
\caption{Mean $p_t$ (left) and its fluctuations (right) vs.\
centrality with data \cite{Adler:2003cb,PHENIX}. Preliminary
$\Lambda$ data is from \cite{StarMean}. Centrality in all figures
is determined by the number of participants relative to the b=0
value.} \label{fig:PHENIX}
\end{center}
\end{figure}
In \cite{Gavin04} I use the Boltzmann transport equation to show
that thermalization alters the average transverse momentum
following
\begin{equation}\label{eq:meanPt}
    \langle p_t\rangle = \langle p_t\rangle_o S + \langle
    p_t\rangle_e (1-S),
\end{equation}
where $S\equiv e^{-{\mathcal N}}$ is the probability that a
particle escapes the collision volume without scattering. The
initial value $\langle p_t\rangle_{o}$ is determined by the
particle production mechanism. If the number of collisions
$\mathcal{N}$ is small, $S\approx 1 -\mathcal{N}$ implies the
random-walk-like increase of $\langle p_t\rangle$ relative to
$\langle p_t\rangle_{o}$. For a longer-lived system, energy
conservation limits $\langle p_t\rangle$ to a local equilibrium
value $\langle p_t\rangle_e$ fixed by the temperature.

As centrality is increased, the system lifetime increases,
eventually to a point where local equilibrium is reached.
Correspondingly, the survival probability $S$ in (\ref{eq:meanPt})
decreases with increasing centrality. The average $p_t$ peaks for
impact parameters near the point where equilibrium is established.
The behavior in events beyond that centrality depends on how the
subsequent hydrodynamic evolution changes $\langle p_t\rangle_e$
as the lifetime increases. Systems formed in the most central
collisions experience a cooling that reduces (\ref{eq:meanPt})
with proper time $\tau$ as $\langle p_t\rangle_e\propto
\tau^{-\gamma}$ \cite{Gavin04}.

Thermalization can explain the behavior in fig.~\ref{fig:PHENIX}.
The survival probability is $S = \exp\{-\int_{\tau_0}^{\tau_F}
\nu(\tau) d\tau\} \approx (\tau_0/\tau_F)^\alpha$, where $\nu =
\langle \sigma v_{\rm rel}\rangle n(\tau)$ for $\sigma$ the
scattering cross section, $v_{\rm rel}$ the relative velocity, and
$\tau_{0,F}$ the formation and freeze out times. Longitudinal
expansion implies $n(\tau)\propto \tau^{-1}$, yielding the power
law with $\alpha = \nu_0\tau_0$. To fit the measured centrality
dependence, I assume $\alpha = 4$ and $\gamma = 0.15$ in central
collisions, and parameterize $S(N_{\rm part})$ by taking $\alpha
\propto N_{\rm part}^{1/2}$ and $\tau_F -\tau_0 \propto N_{\rm
part}$, where $N_{\rm part}$ is the number of participants. I take
the same $\alpha$ for all species, as appropriate for parton
scattering.

Dynamic fluctuations depend on two-body correlations and,
correspondingly, are quadratic in the survival probability. In
\cite{Gavin04} I add Langevin noise terms to the Boltzmann
equation to describe the fluctuations of the phase space
distribution. A simple limit is obtained when the initial
correlations are independent of those near local equilibrium,
    $\langle \delta p_{t1}\delta p_{t2}\rangle
    =\langle \delta p_{t1}\delta p_{t2}\rangle_o S^2
    + \langle \delta p_{t1}\delta p_{t2}\rangle_e (1-S)^2$;
this form was used in \cite{Gavin04}. Alternatively, if the
initial correlations are not far from the local equilibrium value,
I find
\begin{equation}\label{eq:ptFluct}
    \langle \delta p_{t1}\delta p_{t2}\rangle
    =\langle \delta p_{t1}\delta p_{t2}\rangle_o S^2
    + \langle \delta p_{t1}\delta p_{t2}\rangle_e (1-S^2).
\end{equation}
Here I use (\ref{eq:ptFluct}), which provides somewhat better
agreement with the latest STAR data in the peripheral region where
thermalization is incomplete (and my model assumptions most
applicable). As before, the initial quantity $\langle \delta
p_{t1}\delta p_{t2}\rangle_o$ is determined by the particle
production mechanism, while $\langle \delta p_{t1}\delta
p_{t2}\rangle_e$ describes the system near local equilibrium.

To estimate $\langle \delta p_{t1}\delta p_{t2}\rangle_o$ for
nuclear collisions, I apply the wounded nucleon model to describe
the soft production that dominates $\langle p_t\rangle$ and
$\langle \delta p_{t1}\delta p_{t2}\rangle$, to find
\begin{equation}\label{eq:wnm}
    \langle \delta p_{t1}\delta p_{t2}\rangle_o =
    {{2\langle \delta p_{t1}\delta p_{t2}\rangle_{pp}}\over{N_{\rm part}}}
    \left({{1+R_{pp}}\over{1+R_{AA}}}\right)
\end{equation}
\cite{Gavin04}. The pre-factor is expected because
(\ref{eq:Dynamic}) measures relative fluctuations and, therefore,
should scale as $N_{\rm part}^{-1}$. The term in parentheses
accounts for the normalization of (\ref{eq:Dynamic}) to $\langle
N(N-1)\rangle$; $R_{AA}$ scales as $N_{\rm part}^{-1}$
\cite{PruneauGavinVoloshin}. ISR measurements imply $\langle
\delta p_{t1}\delta p_{t2}\rangle_{pp}/\langle p_t\rangle_{pp}^2
\approx 0.015$. I use {\textsc{HIJING}} to estimate $R_{pp}$ and
$R_{AA}$.

Near local equilibrium, spatial correlations occur because the
fluid is inhomogeneous -- it is more likely to find particles near
a dense clump. These spatial correlations fully determine the
momentum correlations since the distribution at each point is
thermal. The mean $p_t$ at each point is proportional to the
temperature $T(\mathbf{x})$, so that $\langle \delta p_{t1}\delta
p_{t2}\rangle_e\sim \int r(\mathbf{x}_1,\mathbf{x}_2) {\delta
T}(\mathbf{x}_1) {\delta T}(\mathbf{x}_2)$, where
$r(\mathbf{x}_1,\mathbf{x}_2)$ is the spatial correlation
function, $\delta T = T - \overline{T}$, and $\overline{T}$ is a
density-weighted average. I take $n$ ($\propto T^3$) and $r$ to be
Gaussian with the transverse widths $R_t$ and $\xi_t$,
respectively the system radius and correlation length.
In \cite{Gavin04} I obtain
\begin{equation}\label{eq:near}
    \langle \delta p_{t1}\delta p_{t2}\rangle_{e}
    = F{{\langle p_t\rangle^2R_{AA}}\over{ 1+R_{AA}}}
\end{equation}
where $R_{AA}$ is given by (\ref{eq:DynamicMult}) \cite{Gavin04}.
The dimensionless quantity $F$ depends on $\xi_t/R_t$. I compute
$F$ assuming $R_t\propto N_{\rm part}^{1/2}$; $F = 0.046$ for
$\xi_t/R_t = 1/6$. I again use \textsc{HIJING} to estimate
$R_{AA}$. Calculations are compared to PHENIX data in
fig.~\ref{fig:PHENIX} (right), and agree within the large
uncertainties.

\begin{figure}
\begin{center}
\includegraphics[width=3.5in]{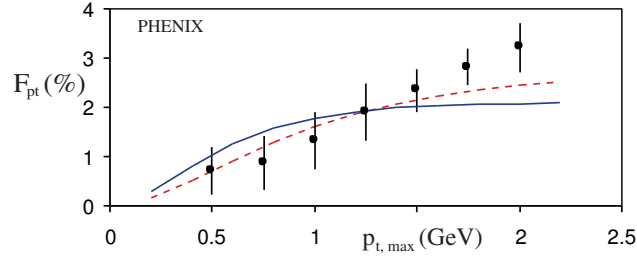}
\caption{Dynamic $p_t$ fluctuations with PHENIX data for central
Au+Au as a function of the acceptance 0.2 GeV~$< p_t < p_{t,\,{\rm
max}}$ \cite{PHENIX}.} \label{fig:ptMax}
\end{center}
\end{figure}
PHENIX also reports $F_{p_t}$ in the acceptance region
$0.2$~GeV~$< p_t < p_{t,\,{\rm max}}$ as a function of
$p_{t,\,{\rm max}}$, with the intent of identifying the role of
jets in producing fluctuations \cite{PHENIX}. To exhibit the
acceptance dependence of $F_{p_t}$, I use (\ref{eq:DynamicMult})
and (\ref{eq:Dynamic}) to estimate $\langle\delta p_{t1}\delta
p_{t2}\rangle \approx \sigma^2R_{AA}$, where $\sigma^2 = \langle
p_t^2\rangle -\langle p_t\rangle^2$. This result is exact for
short range transverse correlations. I then write the observed
quantity $F_{p_t,\,{\rm obs}}\approx (\langle N
\rangle\langle\delta p_{t1}\delta p_{t2}\rangle/2\sigma^2)_{\rm
obs} \propto \langle N\rangle_{\rm obs} R_{AA,\, {\rm obs}}$. The
observed multiplicity grows strongly as $p_{t,\,{\rm max}}$ is
increased, since $\langle N\rangle_{\rm obs}$ is the integral of
the $p_t$ distribution over the acceptance. To determine the
$p_{t,\,{\rm max}}$ dependence of $R_{AA,\,{\rm obs}}$, let the
probability that an individual particle falls in the acceptance
region be $a$. The average number of detected particles is
$\langle N\rangle_{\rm obs} = a\langle N\rangle$. For a binomial
distribution, one has $\langle N^2\rangle_{\rm obs} = a^2\langle
N^2\rangle+a(1-a)\langle N\rangle$, so that (\ref{eq:DynamicMult})
implies $R_{AA,\, {\rm obs}} = R_{AA}$, the value for full
acceptance. It follows that
\begin{equation}\label{eq:ptmax}
F_{p_t,\,{\rm obs}}\approx F_{p_t}{{\langle N\rangle_{\rm
obs}}\over {\langle N\rangle}}
\end{equation}
where $F_{p_t}$ and $\langle N\rangle$ are for full acceptance. I
emphasize that this estimate only includes trivial acceptance
effects. I omit contributions to the $p_{t,\,{\rm max}}$
dependence from the onset of jets, as well as changes in the
correlations due to restricting $p_t$.

\begin{figure}
\begin{center}
\includegraphics[width=2.4in]{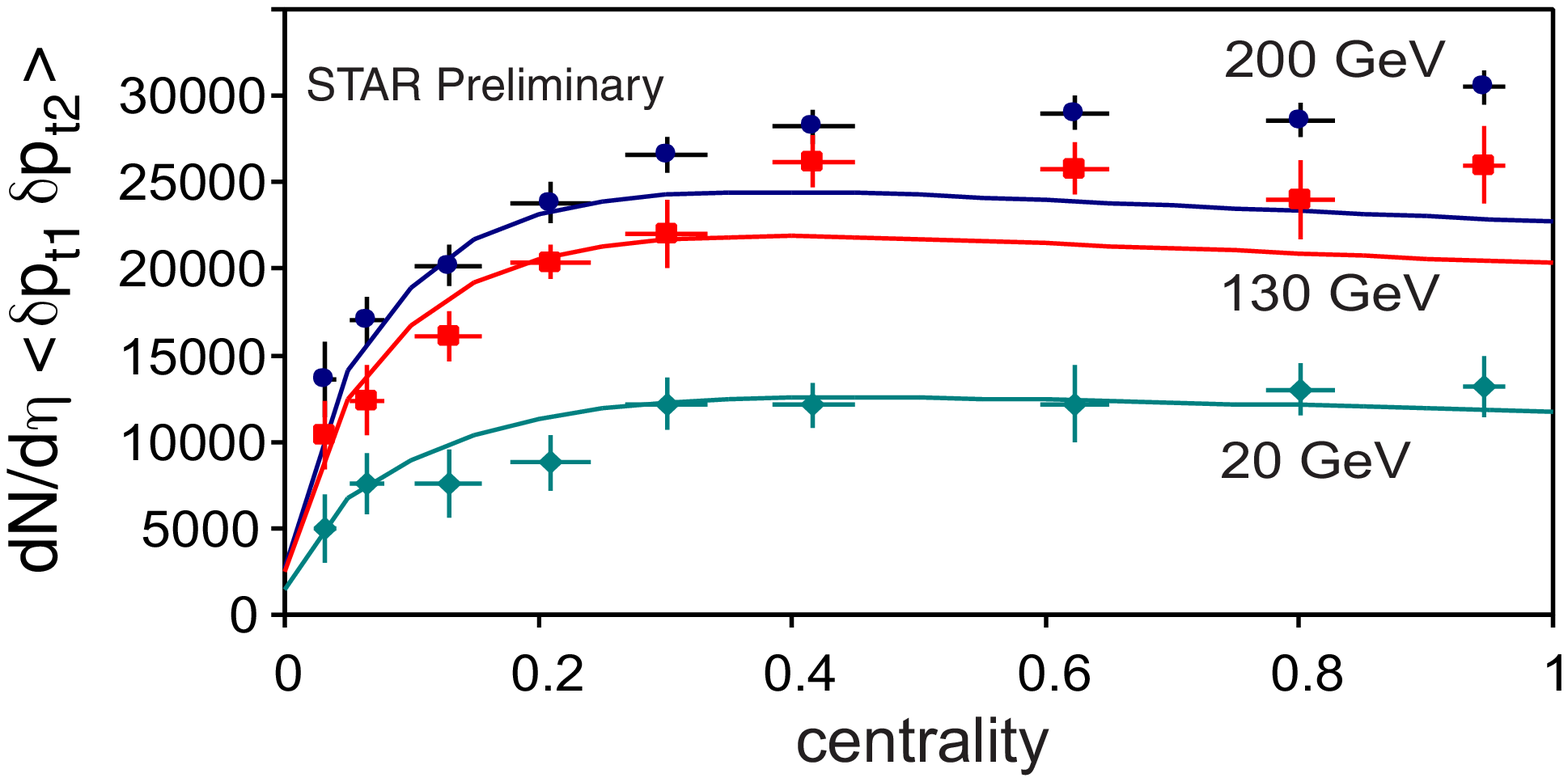}
\includegraphics[width=2.5in]{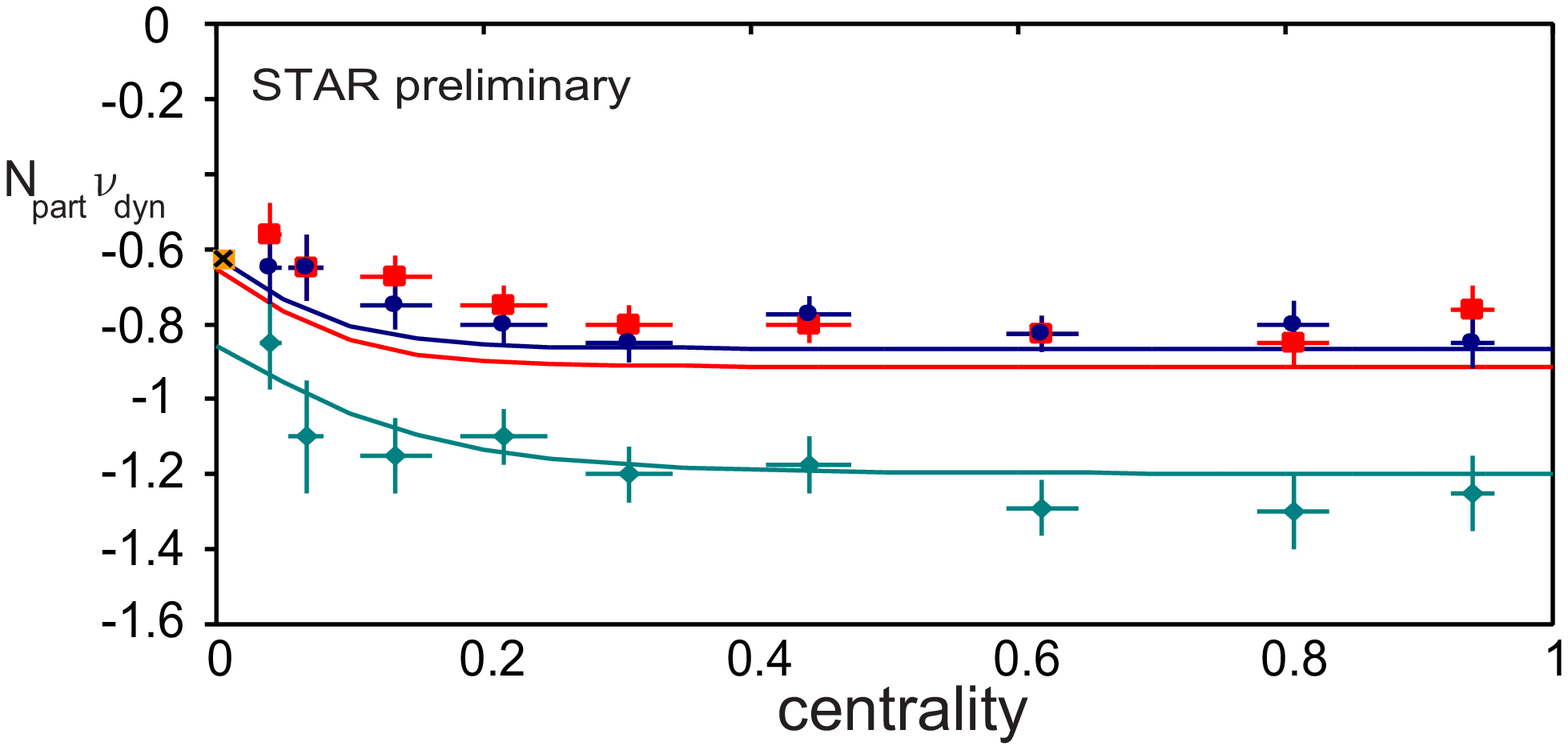}
\caption{Dynamic $p_t$ (left) and net charge (right) fluctuations
with STAR data \cite{Pruneau}. The curves and data are for beam
energies 20, 130, and 200 GeV. The behavior of $\nu_{\rm dyn}$
appears inverted because this quantity is
negative.}\label{fig:ptFluct}
\end{center}
\end{figure}

The estimated $p_{t,\,{\rm max}}$ dependence of $F_{p_t}$ accounts
for the data in fig.~\ref{fig:ptMax}. The solid curve is computed
for $\langle N\rangle_{\rm obs} \propto 1-e^{-p_{t,\,{\rm
max}}/T}(1+{{p_{t,\,{\rm max}}}/{T}})$ for a thermal distribution
at the freeze out temperature obtained from the measured $\langle
p_t\rangle$; the dashed curve for $T = 600$~MeV is included to
show the $T$ dependence. These curves are arbitrarily normalized,
since my computed values in fig.~\ref{fig:PHENIX} fall above the
most central PHENIX data (but below STAR, see
fig.~\ref{fig:ptFluct}). Jets may account for the discrepancy at
the highest $p_{t,\,{\rm max}}$.

The agreement in figs.~\ref{fig:PHENIX} and \ref{fig:ptMax} is
excellent -- and therefore somewhat puzzling -- since I have
neglected collective flow \cite{Voloshin}. Flow must contribute to
fluctuations at some level. However, thermalization and radial
flow are manifestations of multiple scattering in different
regimes; the flow velocity is uniquely defined only when local
equilibrium is established. Both effects enhance $p_t$, so that
the flow contribution may be partly concealed by my parameter
choices.

Preliminary STAR $p_t$ and net charge fluctuation data in
fig.~\ref{fig:ptFluct} show fluctuations at 20, 130 and 200 GeV
beam energies. To compute this energy dependence, I assume $\alpha
\propto N$, $R_{AA}\propto N^{-1}$ and take $N$ from data. The
$p_t$ fluctuation data show a tantalizing similarity to the
calculations, although the experimental uncertainty must be
reduced to firmly establish the behavior at high multiplicity.
Net charge fluctuations in fig.~\ref{fig:ptFluct} (right) are
characterized by $\nu_{dyn} = R^{++} + R^{--} -2R^{+-}$ for
$R^{ab}$ given by (\ref{eq:DynamicMult}) \cite{Pruneau2}. I
therefore expect $\nu_{dyn}$ to satisfy $\nu_{dyn} =\nu_o S^2 +
\nu_e (1-S^2)$. I use pp data to fix $\nu_0$ and compute
$\nu_e/\nu_0$ assuming the longitudinal narrowing of the
correlation length seen in balance function data \cite{Westfall2}.
The agreement of in-progress calculations with preliminary data
\cite{Pruneau,Pruneau2} is encouraging.

Deviations of the calculated fluctuations from data in central
collisions at the highest energies are expected due to flow
\cite{Voloshin} and jets \cite{Mitchell}. Such deviations may be
seen in fig.~\ref{fig:ptFluct}. Nevertheless, I stress that hard
scattering effects -- including jet quenching -- are incorporated
in my calculations via the {\textsc{HIJING}} $R_{AA}$ in
(\ref{eq:near}). {\textsc{HIJING}} alone does not describe the
data discussed here \cite{Westfall2}.

Thermalization is consistent with the acceptance dependence in
fig.~\ref{fig:ptMax}. I surmise that (\ref{eq:ptmax}) holds for a
range of models, since my argument was rather general. That said,
precise measurements may yet reveal jet and flow contributions
\cite{Mitchell,Voloshin}. I find that $\nu_{\rm dyn}$ and
$\langle\delta p_{t1}\delta p_{t2}\rangle/\langle p_t\rangle^2$
minimize trivial acceptance effects and are correspondingly less
ambiguous.

\section*{Acknowledgements}

I thank M.~Abdel~Aziz, J.~Mitchell, C.~Pruneau, M.~Tannenbaum, and
G.~Westfall for discussions. This work was supported in part by a
U.S. National Science foundation CAREER award under grant
PHY-0348559.


\end{document}